\documentclass[sigconf]{acmart}

\AtBeginDocument{%
  }

\copyrightyear{2025}
\acmYear{2025}
\setcopyright{rightsretained}
\acmConference[MobileHCI '25 Adjunct]{Adjunct Proceedings of the 27th International Conference on Mobile Human-Computer Interaction}{September 22--25, 2025}{Sharm El-Sheikh, Egypt}

\newcommand{\meansd}[2]{$M = #1,\: SD = #2$}

\newcommand{\anova}[4]{$F_{#1,#2}=#3$, \pval{#4}} % Classic ANOVA: DF, DFDen, F and p
\newcommand{\anovawefp}[5]{$F_{#1,#2}=#3$, \pval{#4}, \effp{#5}}

\newcommand{\effp}[1]{$\eta^{2}_{p}=#1$} % ANOVA with partial effect
\newcommand{\effd}[1]{$d=#1$}

\newcommand{\pval}[1]{$p#1$}
\newcommand{\pbonf}[1]{$p_{bonf}#1$}

\usepackage{tcolorbox}

\newcommand{\Pilotone}[1]{\textit{Pilot 1#1}}
\newcommand{\Studyone}[1]{\textit{Study 1#1}}
\newcommand{\Pilottwo}[1]{\textit{Pilot 2#1}}
\newcommand{\Pilotthree}[1]{\textit{Pilot 3#1}}
\newcommand{\Pilotfour}[1]{\textit{Pilot 4#1}}
\newcommand{\PilotAll}[1]{\textit{Pilot 1-4}}

\newcommand{\Progressive}[1]{\textit{Progressive{#1}}}
\newcommand{\Full}[1]{\textit{Full{#1}}}
\newcommand{\Word}[1]{\textit{Word{#1}}}

\newcommand{\Mobility}[1]{\textit{Mobility{#1}}}
\newcommand{\Sitting}[1]{\textit{Sitting{#1}}}
\newcommand{\Walking}[1]{\textit{Walking{#1}}}

\newcommand{\WordGap}[1]{\textit{Word-Gap{#1}}}
\newcommand{\NoGap}[1]{\textit{NoGap{#1}}}
\newcommand{\WithGap}[1]{\textit{Gap{#1}}}

\newcommand{\Recall}[1]{\textit{{#1}Recall}}
\newcommand{\DifficultToUnderstand}[1]{\textit{{#1}DifficultToUnderstand}}
\newcommand{\Confusing}[1]{\textit{{#1}Confusing}}
\newcommand{\EasyToRemember}[1]{\textit{{#1}EasyToRemember}}
\newcommand{\AbsorbedInLearning}[1]{\textit{{#1}AbsorbedInLearning}}
\newcommand{\Enjoyable}[1]{\textit{{#1}Enjoyable}}
\newcommand{\Preference}[1]{\textit{Preference{#1}}}

\renewcommand{\quote}[1]{``\textit{#1}''}

\usepackage{subcaption}
\usepackage{multirow}
\usepackage{xcolor}
\usepackage{enumitem}

\usepackage[nomessages]{fp}% http://ctan.org/pkg/fp (math)
\usepackage{soul} % text highlighting
\usepackage{pifont}
\usepackage{longtable}
\usepackage{makecell}

\usepackage{arydshln} % add dash lines

\usepackage{hyperref} % cross-reference with custom text

\newcommand{\ha}[1]{\colorbox{orange!30}{\strut #1}} % lighter orange
\newcommand{\hb}[1]{\colorbox{green!30}{\strut #1}} % lighter yellow

\newlength\maxlen
\def\databarlength{xx.xx} % 4 digits
\begin{document}

%%
%% The "title" command has an optional parameter,
%% allowing the author to define a "short title" to be used in page headers.
\title[Progressive Sentences]{Progressive Sentences: Combining the Benefits of Word and Sentence Learning}

%%
%% The "author" command and its associated commands are used to define
%% the authors and their affiliations.
%% Of note is the shared affiliation of the first two authors, and the
%% "authornote" and "authornotemark" commands
%% used to denote shared contribution to the research.

\author{Nuwan Janaka}
\email{nuwanj@u.nus.edu}
\orcid{0000-0003-2983-6808}

\affiliation{%
  \department{Smart Systems Institute; Synteraction Lab}
  \institution{National University of Singapore}
%   \city{Singapore}
  \country{Singapore}
}

\author{Shengdong Zhao}
\authornote{Corresponding Author.}
\email{shengdong.zhao@cityu.edu.hk}
\orcid{0000-0001-7971-3107}

\affiliation{%
\department{Synteraction Lab, School of Creative Media \& Department of Computer Science}
\institution{City University of Hong Kong}
\city{Hong Kong}
  \country{China}
}

\author{Ashwin Ram}
\authornote{Research conducted while at Synteraction Lab, University of Singapore.}
\email{ashwinram10@gmail.com}
\orcid{0000-0003-1430-8770}

\affiliation{%
  \department{HCI Lab, Saarland Informatics Campus}
  \institution{Saarland University}
  \city{Saarbrücken}
  \country{Germany}
}

\author{Ruoxin Sun}
\email{sun.ruoxin@u.nus.edu}
\orcid{0009-0000-6248-2824}

\affiliation{%
    \department{School of Computing; Synteraction Lab}
    \institution{National University of Singapore} 
    % \city{Singapore}
    \country{Singapore}
}

\author{Sherisse Tan Jing Wen}
\email{sherisse_tjw@u.nus.edu}
\orcid{0009-0003-7078-2642}

\affiliation{%
  \department{School of Computing}
    \institution{National University of Singapore} 
    % \city{Singapore}
  \country{Singapore}
}

\author{Danae Li}
\email{amethyst_li@outlook.com}
\orcid{0009-0009-9974-6796}

\affiliation{%
    \department{MusicX Lab}
    \institution{SKYWORK AI PTE. LTD.}
    \city{Beijing}
    \country{China}
}

\author{David Hsu}
\email{dyhsu@comp.nus.edu.sg}
\orcid{0000-0002-2309-4535}

\affiliation{%
 \department{School of Computing; Smart Systems Institute}
  \institution{National University of Singapore}
  % \city{Singapore}
  \country{Singapore}
}

%%
%% By default, the full list of authors will be used in the page
%% headers. Often, this list is too long, and will overlap
%% other information printed in the page headers. This command allows
%% the author to define a more concise list
%% of authors' names for this purpose.
\renewcommand{\shortauthors}{Janaka et al.}

%%
%% The abstract is a short summary of the work to be presented in the
%% article.
% one sentence motivation
% what was done
% what was found (specific)

\begin{abstract}

\label{sec:abstract}
The rapid evolution of lightweight consumer augmented reality (AR) smart glasses (a.k.a. optical see-through head-mounted displays) offers novel opportunities for learning, particularly through their unique capability to deliver multimodal information in just-in-time, micro-learning scenarios. 
This research investigates how such devices can support mobile second-language acquisition by presenting progressive sentence structures in multimodal formats. In contrast to the commonly used vocabulary (i.e., word) learning approach for novice learners, we present a ``progressive presentation'' method that combines both word and sentence learning by sequentially displaying sentence components (subject, verb, object) while retaining prior context. Pilot and formal studies revealed that progressive presentation enhances recall, particularly in mobile scenarios such as walking. Additionally, incorporating timed gaps between word presentations further improved learning effectiveness under multitasking conditions. 
Our findings demonstrate the utility of progressive presentation and provide usage guidelines for educational applications---even during brief, on-the-go learning moments.

\end{abstract}

%%
%% The code below is generated by the tool at http://dl.acm.org/ccs.cfm.
%% Please copy and paste the code instead of the example below.
%%


%%y
%% Keywords. The author(s) should pick words that accurately describe
%% the work being presented. Separate the keywords with commas.

%%
%% This command processes the author and affiliation and title
%% information and builds the first part of the formatted document.
\maketitle

% INTRODUCTION (deliver motivation, a bit into second page)
% State of the World
% The big BUT .... (stated as what matters to people)
% Therefore, we did....
% The key findings are...
% The contributions of the work are... (1 to 3)

\section{Introduction}

\label{sec:introduction}

The rise of mobile learning reflects growing demand for educational tools that accommodate fast-paced, multitasking lifestyles---particularly in language learning, where consistent practice is essential \cite{shail_using_2019, cai_waitsuite_2017}. However, many learners struggle to maintain regular study due to frequent interruptions, cognitive load fluctuations, and physical limitations such as small mobile screens \cite{zhao_stationary_2018, curum_cognitive_2021, park_effects_2018}. These challenges call for lightweight, cognitively efficient solutions that integrate seamlessly into daily routines.

\begin{figure*}
  \includegraphics[width=\textwidth]{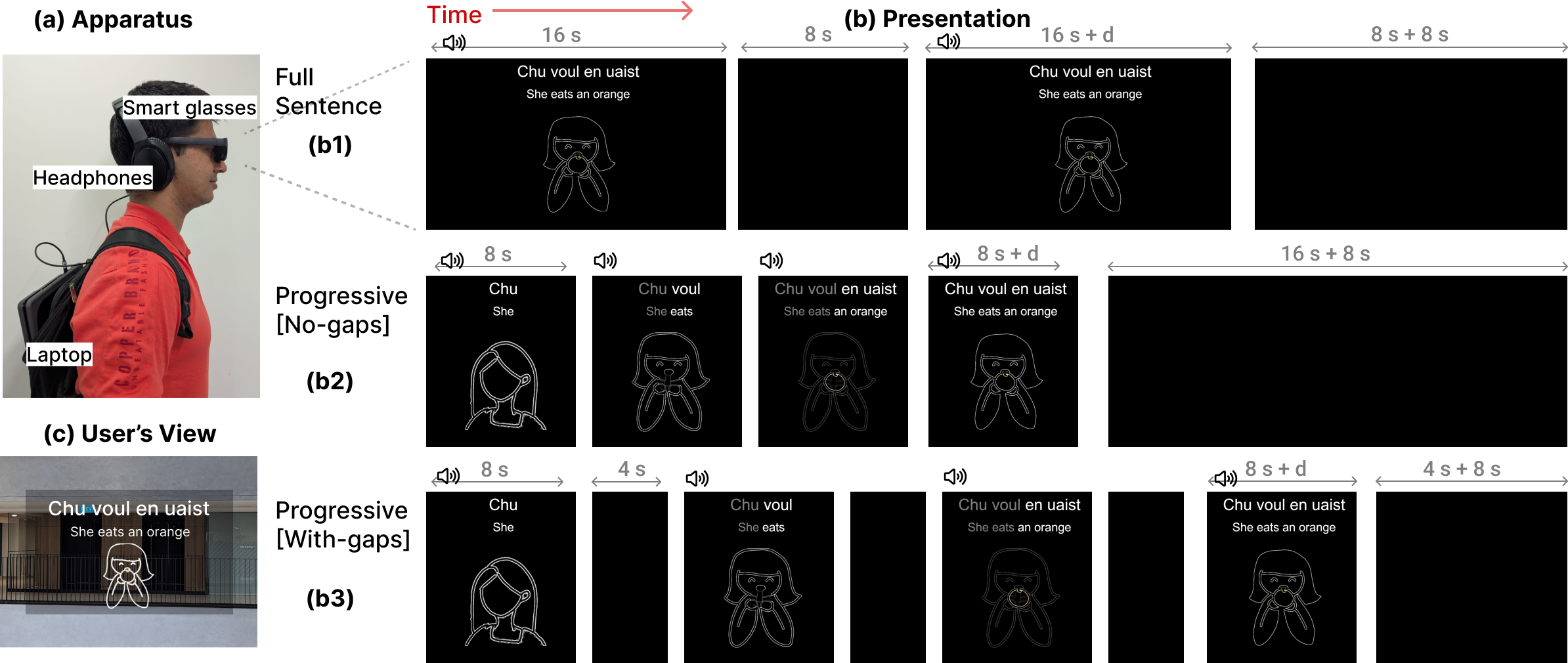}
 \caption{Apparatus and presentation formats used in the study. 
  (a) A participant wearing AR smart glasses and headphones while walking. 
  (b) The participant views learning content through the smart glasses, illustrating the \Full{} (b1) and \Progressive{} presentation formats, including the \NoGap{} (b2) and \WithGap{} (b3) conditions. 
  The durations for each modality (L1/L2 text, L1/L2 audio, images) and no-content gaps were kept consistent. An extra 8 seconds was added between sentences, and an additional \(d = 3\) seconds was included for the final full sentence to ensure adequate viewing time. 
  Black regions in the smart glasses display represent \textit{transparent} areas. 
  (c) Simulated user's view showing how digital content is superimposed on the physical environment. Due to the tinting of smart glasses, the content appears darker than normal. In reality, users do not notice this effect as the frame of the smart glasses blocks peripheral vision.}
  % \Description{
  %   The figure shows three components. 
  %   (a) A side view of a participant wearing AR smart glasses and headphones connected to a backpack laptop. 
  %   (b) Three timelines illustrating content presentation methods: 
  %     (b1) Full sentence format, where the sentence "Chu voul en uaist / She eats an orange" is presented for 16 seconds, followed by 8 seconds of blank screen. 
  %     (b2) Progressive No-gaps format, where the sentence appears in parts ("Chu", "Chu voul", then full sentence), with each part building upon the previous without any blank interval.
  %     (b3) Progressive With-gaps format, where the parts appear sequentially ("Chu", blank, "Chu voul", blank, full sentence), each separated by a gap. 
  %   All formats use illustrations and audio playback for both the L2 language and L1 translation. 
  %   (c) A simulated view of what the user sees through the smart glasses, showing digital text and imagery overlaid on a real-world scene.
  % }
  \label{fig:apparatus_presentation}
\end{figure*}

Advances in digital technology offer promising avenues to meet these needs. In particular, consumer-grade augmented reality (AR) devices—lightweight Optical See-Through Head-Mounted Displays (OST-HMDs or AR smart glasses)—are emerging as viable tools for everyday use \cite{itoh_towards_2021, azuma_road_2019}. These assisted reality devices, such as the XReal Air, Rokid, and TCL RayNeo 2, overlay digital content in the user's field of view, enabling quick-glance, context-independent interactions \cite{rauschnabel_what_2022}. Unlike immersive AR systems that spatially anchor content, these streamlined devices support subtle, ambient engagement, making them well-suited for opportunistic mobile learning \cite{janaka_visual_2022}.

Building on this potential, we investigate how lightweight AR glasses can support mobile language learning. A key question arises: \textit{how can multimodal information be composed and presented effectively within the constraints of these devices?} Rather than developing complex, context-aware applications, we focus on optimizing content presentation to suit device limitations. This approach addresses both technical constraints of OST-HMDs and real-world usage—typically brief, multitasked interactions.

Grounded in Cognitive Theory of Multimedia Learning \cite{mayer_cognitive_2014, mayer_multimedia_2002}, our study examines how combinations of visual (e.g., text, images) and auditory (e.g., pronunciation) elements enhance language learning while managing cognitive load. We aim to develop presentation methods that are lightweight yet effective, improving outcomes within the limits of current OST-HMDs.

To this end, we conducted pilot and formal experiments evaluating sentence presentation techniques using lightweight OST-HMDs. We tested different multimodal combinations—text, audio, images—to assess their effect on learning. We also examined how multitasking influences learning, acknowledging varied usage contexts from stationary to mobile settings.

Our findings show that a progressive sentence presentation strategy supports language learning during mobile multitasking. This method incrementally introduces sentence components before presenting the full sentence, helping learners grasp both individual elements and overall structure. Additionally, inserting strategic word gaps enhances recall in mobile contexts.

This work provides preliminary empirical evidence on AR-assisted language learning and offers design insights for educational applications. Our findings highlight the potential of lightweight OST-HMDs to support natural, engaging information acquisition.

\section{Related Work}
\label{sec:related_work}

\subsection{Mobile Language Learning and Augmented Reality}
Mobile microlearning leverages short, intermittent periods to deliver bite-sized language content (e.g., vocabulary), supporting acquisition during everyday tasks \cite{shail_using_2019, cai_waitsuite_2017}. It has gained traction due to its accessibility and convenience \cite{edge_micromandarin_2011, dingler_language_2017, edge_memreflex_2012}. To enhance retention, prior work has explored situated contexts, such as location-based microlearning \cite{edge_micromandarin_2011}. Recently, OST-HMDs or AR smart glasses have emerged as promising platforms for context-aware learning, offering hands-free, heads-up access to digital content superimposed in the user's view (i.e., physical world) \cite{itoh_towards_2021}. Systems like \textit{ARbis Pictus} \cite{ibrahim_arbis_2018} and \textit{VocabulARy} \cite{weerasinghe_vocabulary_2022} improve recall over non-AR approaches (e.g., desktop flashcards or tablet annotations) by embedding visual annotations in situated physical contexts. While most prior work has focused on vocabulary learning in both AR and non-AR settings, emerging evidence suggests that embedding vocabulary within broader semantic structures—such as sentence-level presentations or narratives—enhances comprehension and retention in stationary, non-AR environments \cite{leong_putting_2024, chen_retassist_2024, teng_effectiveness_2023}. Our work extends this by introducing a progressive sentence disclosure method on AR smart glasses, integrating isolated vocabulary with contextual sentence presentation under mobile multitasking conditions.

\subsection{Multimodal Presentation and Cognitive Load}
Mayer's Cognitive Theory of Multimedia Learning (CTML), along with the Multimedia Principle, posits that effective learning combines visual, auditory, and textual modalities to support richer encoding and improved retention \cite{mayer_multimedia_2002, mayer_cognitive_2014}. However, poorly designed multimodal presentations can increase cognitive load, especially for novice learners \cite{eng_keep_2020, torcasio_use_2010}. This aligns with Sweller's Cognitive Load Theory, which distinguishes between \textit{intrinsic load} (material complexity), \textit{extraneous load} (load from poor design), and \textit{germane load} (resources allocated to schema construction) \cite{torcasio_use_2010, mayer_cognitive_2014}. These loads interact dynamically: reducing extraneous load frees cognitive capacity for germane processing. In language learning, intrinsic load is high due to unfamiliar vocabulary and grammar, making it essential to minimize extraneous load through effective presentations/interfaces while promoting germane load. Strategies such as the segmenting principle—breaking content into smaller, progressive units—help manage extraneous load by enabling incremental processing \cite{lavie_load_2004, mayer_cognitive_2014}. Our approach builds on these principles by employing progressive disclosure: sequentially revealing sentence components with synchronized multimodal cues (text, images, audio). This design reduces extraneous load and supports multitasked learning \cite{ram_lsvp_2021, janaka_visual_2022}, while potentially enhancing germane load by reinforcing semantic links between words in context.

\section{Study Overview}

This research investigates effective sentence presentation styles for language learning in mobile contexts using assisted reality \cite{rauschnabel_what_2022}. While word-level learning supports beginners \cite{gu_vocabulary_1996}, sentence-based approaches provide richer context for deeper learning \cite{leong_putting_2024, noauthor_introducing_2023, chen_retassist_2024}. To explore this, we conducted informal pilot studies to refine 1) modalities for sentence presentation (\Pilotone{}, N=5), and 2) chunking styles that support mobile learning (\Pilottwo{}, N=6). These led to a progressive sentence disclosure method that decomposes sentences into subject-verb-object structures. While promising, the pilots also revealed the potential influence of timing gaps between word displays (\Pilotthree{}, N=2), prompting a formal investigation (\Studyone{}, N=12). A follow-up pilot (\Pilotfour{}, N=4) evaluated the method's ecological validity.

\section{Common Setting}
\label{sec:common_setting}

All pilot (\PilotAll{}) and formal (\Studyone{}) studies were conducted in a controlled lab environment to minimize confounding factors (e.g., noise, lighting) and shared the following common elements.

\subsection{Participants}
Participants (total of 27) were adult learners (age 19–26) from the university community, self-reporting full professional fluency in English as their first language (L1) and willingness to learn a second language (L2). Demographics, language background, and fluency are detailed in each study section. All had normal or corrected-to-normal vision and hearing, with no reported impairments.

All procedures were IRB-approved. Participants gave informed consent and were compensated at \$7.25/hour. \textbf{No} participant took part in more than one study.

\subsection{Apparatus}
As shown in Figure~\ref{fig:apparatus_presentation}(a), participants wore XREAL Air glasses (1920×1080 px, 46\textdegree{} FoV, 60Hz) and Bose QuietComfort Bluetooth noise-canceling headphones for audio and to secure the glasses. The AR glasses mirrored a laptop's display via USB\footnote{XREAL Air projects a 130-inch equivalent screen at ~4 meters}. The learning content presentations were controlled using a Python backend and VueJS frontend on a MacBook Pro (14-inch, M3 Pro). During walking conditions, the laptop was carried in a light backpack.
The laptop also shared its screen via Zoom with the experimenter for monitoring. See \url{https://github.com/Synteraction-Lab/ProgressiveSentences} for implementation details.

\subsection{Task}
We simulated two everyday AR usage scenarios:

\textbf{\Walking{} [Multitasked Learning]}: Mobile multitasking scenario simulating real-world contexts like commuting, where users' attention is divided between non-learning (e.g., walking, checking the surroundings) and learning tasks \cite{zhao_stationary_2018}. Participants assembled a 6-piece Duplo Lego block to match a color/shape sample, transporting pieces along a 5-meter path, simulating hands-busy multitasking. Block colors varied to ensure consistent difficulty.

\textbf{\Sitting{} [Focused Learning]}: Seated language learning to simulate stationary usage (e.g., focused learning while sitting).

\subsection{Materials}
Participants studied L2 words or sentences with L1 meanings, depending on the condition.

\subsubsection{L2 Language}
We used Wuggy pseudowords \cite{keuleers_wuggy_2010} as L2, commonly applied in HCI and linguistics (e.g., \cite{sheshadri_learn_2020, macedonia_body_2011}). This controlled for prior knowledge, avoided cognates, and ensured consistent complexity \cite{carter_vocabulary_1998}.

\subsubsection{L1 Sentences}
Focusing on A1 fluency (beginner level) \cite{noauthor_cefr_nodate}, we created 40 English (L1) statements using basic Subject-Verb-Object grammar. Nouns were drawn from Ogden's Basic English \cite{ralli_review_1991}, with common pronouns and function words (articles/prepositions). Each sentence had 2–3 unique L1 words (with 4-5 total words).

\subsubsection{Sentence List}
Each L1 word was mapped to a Wuggy L2 pseudoword. Three co-authors reviewed all pairs to ensure no cognates. L2 sentences were created via word-to-word mappings (see Table~\ref{tab:l1_l2_examples}).

\begin{table*}[hptb]
\caption{Sample L1-L2 words and L1-L2 sentence list}
% \Description{The table contains sample word and sentence pairs across three languages: English (L1), a constructed language called Wuggy (L2), and Norwegian. The left side shows English words and their Wuggy equivalents. The right side presents English sentences with corresponding translations in Wuggy and Norwegian. For example, the English word “he” is translated as “sa” in Wuggy, and the sentence “He reads a book” is rendered as “Sa soan en beal” in Wuggy and “Han leser en bok” in Norwegian.}
\label{tab:l1_l2_examples}
\scalebox{1}{
\begin{tabular}{@{}llllll@{}}
\cmidrule(r){1-2} \cmidrule(l){4-6}
L1 Word (English) & L2 word (Wuggy) &  & L1 Sentence (English) & L2 Sentence (Wuggy) & L2 Sentence (Norwegian)\\ \cmidrule(r){1-2} \cmidrule(l){4-6} 
he & sa &  & He reads a book & Sa soan en beal & Han leser en bok \\
she & chu &  & She eats an orange & Chu voul en uaist & Hun spiser en appelsin \\
reads & soan &  & The boy kicks a ball & Snu bist yills en bune & Gutten sparker en ball \\
\cmidrule(r){1-2} \cmidrule(l){4-6} 
\end{tabular}
}
\end{table*}

\subsubsection{Multimodal Representations}
Audio (L1 and L2 pronunciations) was generated via OpenAI TTS API (tts-1, voice: nova, speed: 0.9) and manually verified. Mismatched L2 phrases were corrected by concatenating word-level audio. Audio duration ranged from 0.1–3 seconds.

Google Image Search was used to retrieve outline-style monochrome icons for each word, phrase, and sentence, minimizing visual distraction \cite{eng_keep_2020, tan_audioxtend_2024}. Three co-authors iteratively rated and refined images for adherence to the selected `Seven C' principles (concise, concrete, coherent, comprehensible, correspondent) \cite{goldman_role_1998}.

\subsubsection{Presentation Layout and Style}
As shown in Figure~\ref{fig:apparatus_presentation}(b), L2 text, L1 text, and images appeared top-center, minimizing visual obstruction. Transparent-outline images preserved central visibility and enabled quick recognition \cite{janaka2023icons, zhou_not_2023}. All visuals were rendered in white for consistency \cite{tan_audioxtend_2024}.

Following prior work \cite{sheshadri_learn_2020, gu_vocabulary_1996} and informal testing, L2 text appeared before its L1 meaning. Audio was synchronized with text; images appeared with L2 text. Words were displayed for at least 6 seconds, phrases (consisting of 2 or more words) for 8 seconds, and sentences for 12 seconds. Each item was shown twice to enable review. We avoided spaced repetition \cite{edge_memreflex_2012} to equalize exposure time. All modalities and gaps were time-matched across conditions (i.e., equal total exposure time). Refer to individual study sections for presentation timing details.

\subsection{Study Design}
All studies (pilot and formal) employed a within-subject, repeated-measures design. Experimental conditions were counterbalanced using a Latin square. To control learning material complexity, 3–4 sentences were grouped per condition with matched sentence lengths and L2 word counts. Learning material groups were presented in a fixed order, while experimental conditions were counterbalanced, to minimize biases across groups.

\subsection{Measures}
Table~\ref{tab:measures} lists all common measures.
We evaluated `remembering' and `understanding' based on Bloom’s Revised Taxonomy \cite{krathwohl_revision_2002}.
Primary metrics were free recall (L2–L1 word pairs, no cues) and cued recall (L2 to L1; \Recall{}) \cite{eng_keep_2020, krathwohl_revision_2002}. \Recall{Word-} and \Recall{Seen-Sentence-} assessed \textit{remembering} for studied content, while \Recall{Unseen-Sentence-} tested \textit{understanding} via novel L2 sentences combining known words \cite{torcasio_use_2010}. For example, if a participant saw `She eats an orange' and `The boy kicks a ball,' an unseen sentence might be `The boy eats a ball,' which, although potentially unrealistic, would be grammatically correct. 
A questionnaire collected all four \Recall{} types in this order: \Recall{Free-}, \Recall{Word-}, \Recall{Seen-Sentence-}, \Recall{Unseen-Sentence-}. Scoring followed prior work \cite{weerasinghe_vocabulary_2022, ibrahim_arbis_2018}, allowing partial credit for correct structure and meaning (see Appendix~\ref{sec:common_setting:scoring}).

\begin{table*}[tbhp]
\caption{Measures used in each study}
% \Description{This table presents the different measures used in the study, categorized into Recall, Cognitive Load, and Preference. Recall includes four sub-measures: Free Recall, Word Recall, Seen-Sentence Recall, and Unseen-Sentence Recall, each measuring how well participants could recall words or sentences in L1 and L2 under various conditions. Cognitive Load includes five Likert-style statements assessing participants’ perceptions of the learning experience, such as difficulty, confusion, memorability, engagement, and enjoyment. Preference is represented by a single measure capturing participants’ ranking of different learning styles.}
\label{tab:measures}
\scalebox{1}{
\begin{tabular}{@{}lll@{}}
\toprule
Measure & Sub-measure & Definition/Operationalization \\ \midrule
Recall 
 & \Recall{Free-} & The total score for writing the correct L1-L2 words pairs without any cues\\
 & \Recall{Word-} & The total score for writing the correct L1 word for the seen (given) L2 word \\
 & \Recall{Seen-Sentence-} & The total score for writing the correct L1 sentence for the seen L2 sentence \\
 & \Recall{Unseen-Sentence-} & The total score for writing the correct L1 sentence for the unseen L2 sentence \\ 
\hline
Cognitive Load & \DifficultToUnderstand{} & \quote{I find the learning content during this session/style difficult to understand.} \\
(Learning \cite{leppink_development_2013, chin_applying_2024}) & \Confusing{} & \quote{The learning content presented by this learning session/style is confusing for me.} \\
 & \EasyToRemember{} & \quote{This learning session/style allows me to easily remember most of the learning content.} \\
 & \AbsorbedInLearning{} & \quote{I was absorbed in using this session/style to learn the new language.} \\
 & \Enjoyable{} & \quote{It is enjoyable to learn a new language with this session/style.} \\ 
\hline
Preference & \Preference{} & The overall preference ranking \\ 
\bottomrule
\end{tabular}
}
\end{table*}

Learning cognitive load ratings were collected via 7-point Likert scales per condition, based on validated measures \cite{leppink_development_2013, chin_applying_2024}.

We also collected preference rankings (\Preference{}) and qualitative feedback through semi-structured interviews on strategies, challenges, and suggestions.

\paragraph{Analysis} 
Data were analyzed using repeated-measures or factorial ANOVA. If assumptions were violated, we used Friedman tests or ART-based ANOVA \cite{wobbrock_aligned_2011}. Normality and sphericity were checked using Shapiro-Wilk and Mauchly’s tests; post-hoc tests used paired t-tests, ART contrasts, or Wilcoxon tests with Bonferroni correction (see Appendix~\ref{sec:common_setting:analysis} for details).

Transcribed interview recordings and open-ended responses in questionnaires were analyzed using thematic analysis as described in Braun and Clarke's methodology \cite{braun_using_2006}.

\subsection{Procedure}
\label{sec:common:procedure}
After informed consent, participants were introduced to tasks, presentation styles, and sample questions. They could practice aloud but not take notes or rehearse outside sessions. Participants completed all conditions in a set order. After each, they completed a questionnaire. Partial answers for immediate recall were allowed, but guessing was discouraged. A 2-minute break followed each condition.

At the end, participants ranked their preferred styles and completed a 5–10 minute semi-structured interview. Sessions lasted around 1 hour, including training. A delayed recall questionnaire was sent online after 7 days to assess retention \cite{ram_lsvp_2021, sheshadri_learn_2020}. Participants were instructed to rely solely on memory.

\section{\Pilotone{}: Identifying Suitable Modalities for Sentences during \Walking{}}

In this pilot, we compared five common modality combinations (i.e., 
\textit{NoImageNoAudio:} L2 Text $\rightarrow$ L1 text; 
\textit{NoImageWithAudio:} L2 text + L2 audio $\rightarrow$ L1 text;
\textit{ImageWithAudio:} L2 text + L2 audio + Image $\rightarrow$ L1 audio;
\textit{ImageWithText:} L2 text + L2 audio + Image $\rightarrow$ L1 text;
\textit{ImageWithTextAudio:} L2 text + L2 audio + Image $\rightarrow$ L1 text + L1 audio
) \cite{mayer_cognitive_2014} with 5 (3F, 2M) participants, to identify complementary modalities for sentence-based learning during \Walking{} (Sec~\ref{sec:common_setting}). 

While there were no statistically significant differences, the text-only condition (i.e., \textit{NoImageNoAudio}) had the lowest recall scores, whereas the text+image+audio combination (i.e., \textit{ImageWithTextAudio}) yielded the highest, aligning with the \textit{Multimedia Principle} \cite{mayer_cognitive_2014}. Specifically, all participants preferred the combination of \textit{L2 text + L2 audio + image, followed by L1 text + L1 audio}. Audio helped with pronunciation (L2) and meaning (L1) without drawing visual attention from the mobile multitasking. However, participants noted difficulty understanding full sentences due to limited support for individual word meanings.

\section{\Pilottwo{}: Evaluating \textit{Progressive Disclosure} during \Walking{}}

\paragraph{\textbf{Progressive Presentation of Sentences}}
To address word-level comprehension within sentences, we developed a \textit{progressive disclosure presentation} (\Progressive{}, Figure~\ref{fig:apparatus_presentation}(b)) based on the \textit{Segmenting Principle}~\cite{mayer_cognitive_2014}. Sentences were broken into subject, verb, and object components, presented sequentially to support stepwise understanding. Previously shown words remained visible, and new ones were highlighted using transparency cues following the \textit{Signaling Principle}~\cite{mayer_cognitive_2014}. Once all components were introduced, the full sentence was displayed.

\paragraph{\Pilottwo{}}
We compared \Progressive{} with full sentences (\Full{}) and individual words presented randomly (\Word{}, no sentence structure), with 6 (4F, 2M) participants during \Walking{}. Although differences were not statistically significant, \Progressive{} yielded the highest recall. All participants preferred it, noting improved memory for both individual words and sentence structure. It also reduced cognitive load compared to \Full{} and helped link L2 and L1 words. 

However, three participants (3/6) suggested adding pauses between words in \Progressive{} to aid memory consolidation.

\section{\Pilotthree{}: Determining Optimal Word Gaps for \Progressive{} during \Walking{}}

This pilot tested gap durations (2s, 4s, 6s, 8s) with two (1F, 1M) participants during \Walking{}. The findings suggested that the gap should be neither too short---making it unnoticeable---nor too long, which could cause participants to forget the connection between words in a sentence. A 4-second gap was identified as a balanced option and selected for further study. While the optimal duration requires further investigation, this finding also aligns with previous research on notification resumption gaps \cite{janaka_notifade_2023}.

% METHOD 
% Quantitative
% Participants
% Apparatus
% Procedure
% Design & Analysis

\section{\Studyone{}: Understanding the Effect of \Mobility{} and \WordGap{} on \Progressive{} Presentation}
\label{sec:study1}

\paragraph{Participants:}
Twelve volunteers (7F, 5M; age \meansd{22.3}{3.3}) participated in the study. Ten were native English speakers; two had professional fluency. All but one had used mobile language apps; two had limited prior AR smart glasses exposure.

\paragraph{Design and Procedure:}
Following Sec~\ref{sec:common:procedure}, the study used a 2×2 design: \Mobility{} (\Sitting{}, \Walking{}) $\times$ \WordGap{} (\NoGap{} = 0s, \WithGap{} = 4s). Participants learned 12 sentences (3 per condition) and 32 new L2 words (8 per condition). 

\begin{figure}[htbp]
  \centering
  \includegraphics[width=1.05\linewidth]{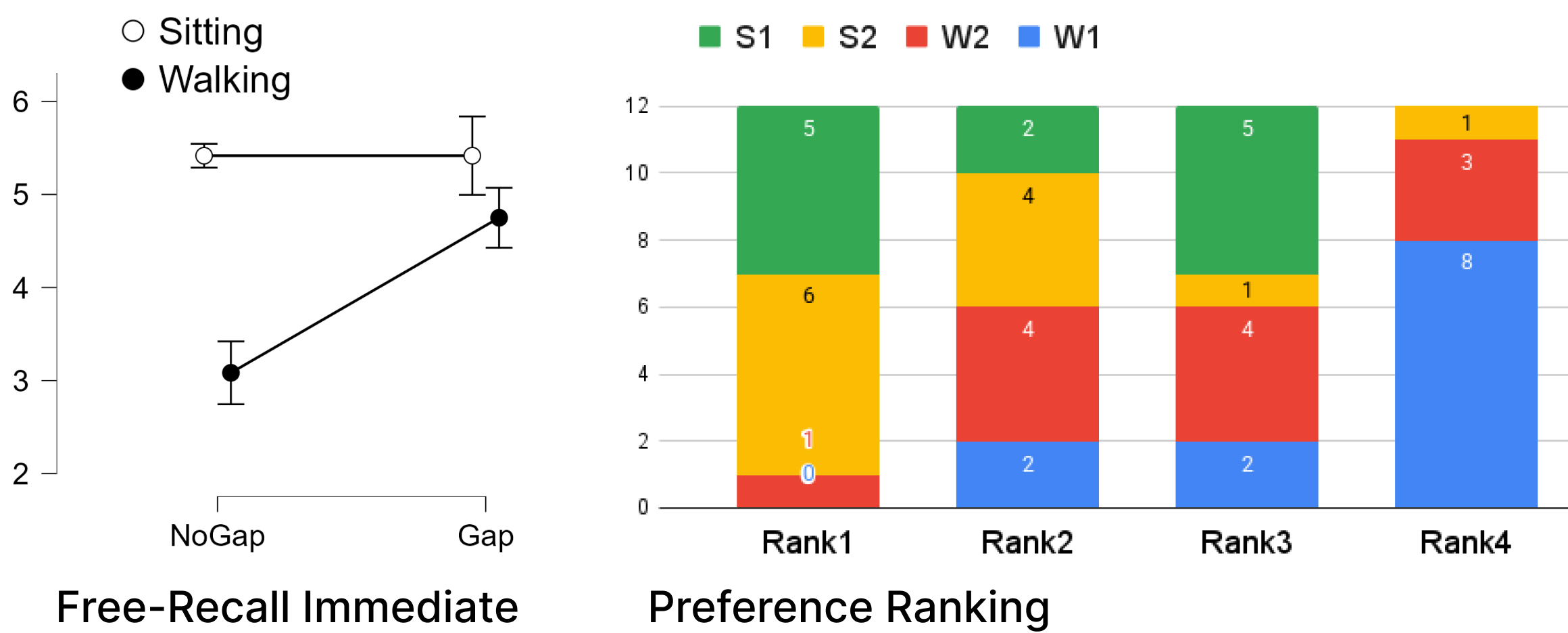}  
  \caption{\Recall{Free-} (Immediate) interactions and \Preference{} rankings for the \Studyone{} (N=12). Here, S = \Sitting{}, W = \Walking{}, 1 = \NoGap{}, and 2 = \WithGap{}. For example, S1 represents, \Sitting{} with \NoGap{}.}
  \label{fig:study1:free_recall_preference}
  % \Description{The figure presents a bar graph comparing Free Recall (Immediate) performance and Preference rankings across four experimental conditions in Study 1. The conditions combine sitting or walking with no-gap or with-gap presentation styles, labeled as S1, S2, W1, and W2. The height of the bars reflects participants' recall scores and ranking preferences for each condition.}
\end{figure}

\paragraph{Results:} Figure~\ref{fig:study1:free_recall_preference} shows the significant results (refer to Appendix~\ref{appendix:study1}: Figure~\ref{fig:study1:measures}, Table~\ref{tab:study1:measures} for details). For \Walking{}, all participants successfully completed the Lego assembly.

\paragraph{\Recall{}:}
Repeated-measures ANOVA with ART \cite{wobbrock_aligned_2011} showed significant main effects of \Mobility{} (\pval{<0.05}).

In \textbf{immediate} \Recall{}, \Sitting{} yielded significantly higher \Recall{Free-}, \Recall{Word-}, and \Recall{Seen-Sentence-} scores (\pval{<0.05}). \Recall{Free-} also showed significant main and interaction effects of \WordGap{} (\anova{1}{33}{5.62}{=0.02}; \anova{1}{33}{5.96}{=0.02}). Post-hoc results (Figure~\ref{fig:study1:free_recall_preference}) revealed \WithGap{} (\meansd{4.9}{2.0}) outperformed \NoGap{} (\meansd{4.3}{1.7}; \pbonf{<0.05}, \effd{0.44}). In \Walking{}, \WithGap{} (\meansd{4.7}{2.2}) significantly exceeded \NoGap{} (\meansd{3.1}{1.8}); no such effect was found in \Sitting{}.

In \textbf{delayed (7-day)} \Recall{}, \Sitting{} again led to significantly better scores for \Recall{Word-} (\anova{1}{33}{13.76}{<0.001}) and \Recall{Seen-Sentence-} (\anova{1}{33}{5.75}{=0.02}). No other significant effects were found.

\paragraph{Learning Cognitive Load:}
All subjective scales showed significant main effects of \Mobility{} (\pval{<0.05}) via ART-based ANOVA. \Sitting{} was rated lower in \DifficultToUnderstand{} and \Confusing{}, and higher in \EasyToRemember{}, \AbsorbedInLearning{}, and \Enjoyable{}. 
Participants attributed this to divided attention during multitasking, which diminished the capacity for remembering learning content.

\paragraph{\Preference{} results (Figure~\ref{fig:study1:free_recall_preference}):} 11/12 preferred \Sitting{} for better focus. One preferred \Walking{} to \quote{keep the mind active}. Word gaps did not affect preferences during \Sitting{}, though two noted increased mind-wandering. During \Walking{}, the majority (9/12) preferred having word gaps as it \quote{gave time to practice and solidify the words in memory} and \quote{one without any gaps is [felt] too fast.} Some of them (6/9) also mentioned that the gaps between words \quote{gave more chances to review previous words and prepare for upcoming words while attention was split.} The rest (3/12) did not notice the gaps.

\paragraph{Subjective Feedback:}
When asked to compare \Progressive{} and \Full{} presentations, most participants (10/12) preferred \Progressive{}, citing its one-to-one word mapping, which helped them understand individual meanings and how words connect. In contrast, \Full{} sentences would require more cognitive effort to link L2 and L1 words, leaving less time for review. Similarly, all participants favored \Progressive{} over random individual words, highlighting the importance of sentence context in supporting comprehension.

% METHOD 
% Quantitative
% Participants
% Apparatus
% Procedure
% Design & Analysis

\section{\Pilotfour{}: Preliminary Ecological Validation of the \Progressive{} Presentation}
\label{sec:pilot4}

To assess generalizability beyond artificial corpora, we used Norwegian (Bokmål) as L2. This pilot followed the \Studyone{} procedure (Sec~\ref{sec:study1}) with differences in materials and a pre-test (see Appendix~\ref{appendix:pilot4}).

Four volunteers (1F, 3M; age \meansd{24.3}{2.1}), without prior knowledge of Norwegian or related languages (e.g., German or Norse), participated in the study. Three were native English speakers; one had professional fluency. All had experience with mobile language apps.

\paragraph{Results:} Each participant learned 12 sentences (3 new sentences per condition) and 28 new vocabulary words (7 new vocabulary words per condition) across four conditions. See details in Appendix~\ref{sec:pilot4:results} (Table~\ref{tab:pilot4:measures}, Figure~\ref{fig:pilot4:measures}).

\Sitting{} resulted in significantly higher \textbf{immediate} \Recall{Free-} than \Walking{} (\anovawefp{1}{3}{32.0}{=0.01}{0.91}); other differences were not significant.
The trends observed were consistent with those in \Studyone{}. Regarding preference and interview feedback, all participants preferred the \Sitting{} over \Walking{} (with no impact of \WordGap{s}), and \WithGap{} over \NoGap{} during \Walking{}. Similar to \Studyone{}, participants reported that word gaps during \Walking{} helped them consolidate memory before the next word appeared, whereas during \Sitting{}, the gaps had minimal perceived impact. 
As one noted, \quote{I felt that in the second walking [without gaps], the words went too fast, but all were ok for sitting.} These findings support the applicability of \Progressive{} for real languages with similar grammar.

% DISCUSSION (what was interesting, what matters)

% What are the implications of your results?
% What do they mean for this topic and your field?
% What is important and worthy of again being called to the reader's attention?
% What was surprising, unexpected, intriguing?
% Did you fulfill the promises you set out in the Intro via the claims of your work?
% Were any hypothesis confirmed or disconfirmed?
% Were the predictions of any theory upheld or refuted?
% What worked and what did not work? And why?

% - add limitations here to guide interpretation of work

\section{Discussion}
\label{sec:general_discussion}

\paragraph{Why is progressive presentation effective for multitasked learning?}

Our findings suggest that \Progressive{} presentation in AR smart glasses—sequentially revealing information while retaining prior content and highlighting new elements—supports language learning by facilitating word familiarity and preserving sentence context. This method aids learners in understanding how words connect meaningfully. These results align with Layered Serial Visual Presentation (LSVP) \cite{ram_lsvp_2021}, which found that persistent, sequential delivery enhances recall in mobile video-learning contexts by improving attention control, consistent with Perceptual Load Theory (PLT) \cite{lavie_load_2004}.

We also found that inserting \textit{word gaps} (i.e., no-content intervals) improves short-term retention during multitasking. Although no significant effects were observed for long-term retention—likely due to small sample size—our findings align with prior work emphasizing system pacing that matches users’ cognitive load and processing capacity \cite{goguey_interaction_2021}. While word gaps had minimal benefit in stationary settings, they were helpful under multitasking, giving learners time to consolidate new information. Although we used a 4-second default, participant feedback suggests that the optimal gap duration may vary based on task demands, highlighting the value of adaptive timing to balance focus between primary and secondary tasks---similar to notification resumption gaps during multitasking\cite{janaka_notifade_2023}.

\paragraph{When and how should progressive presentation be used?}

User feedback indicates that \Progressive{} is most effective as a scaffolding technique for novice learners encountering unfamiliar vocabulary. It enables gradual exposure while maintaining sentence context, supporting both word- and sentence-level comprehension. Once learners develop familiarity, full-sentence presentation may be more efficient, allowing them to infer meanings without stepwise support.

In this study, sentence segmentation and image selection were done manually to ensure coherence. To scale this approach, large language models (LLMs) could automatically segment sentences, while text-to-image models \cite{attygalle_text_image_2025} could generate relevant visuals. Post-processing methods could evaluate image-sentence alignment and assemble semantically coherent sequences \cite{chen_automatic_2024, chen_retassist_2024}.

\subsection{Limitations and Future Work}
\label{sec:limitations_future_work}

While this study offers preliminary evidence, several limitations affect its generalizability. First, the L2 used had grammatical structures similar to L1, limiting insights into structurally different languages. Second, the participant pool was skewed toward tech-savvy individuals. Third, the small sample size and controlled lab settings limit ecological validity. Due to the sample size, no statistically significant effects were observed for long-term recall. Additionally, due to our scoping, the study did not explore context-dependent AR presentations (e.g., 3D-anchored content), and mobile multitasking was not studied independently from walking.

Future work should validate these findings through longitudinal studies involving more diverse populations, broader language sets, and various AR smart glasses in real-world environments. This includes testing under different conditions (e.g., lighting, distractions) to better assess robustness and generalizability. A power analysis using the current results as priors could help determine the minimum required sample size for future studies. Additionally, further research is needed to examine whether \Progressive{} presentation can be effectively extended to other device platforms in multitasking contexts.

% CONCLUSION
% affirm that you have delivered on the claims made in your Introduction
% Summarize the contributions of the work
% Make any key points with which you would like to leave the reader, 
% point to a bright future/better world for your work having been done in it. 

% FUTURE WORK (what are the “big idea” next steps to follow from your work?)
% Avoid merely incremental steps like a todo list

% Try to frame the contributions of the work such that they speak to your broader scholarly community, not just those interested in your narrow topic:

% \section{Conclusion}

%%
%% The acknowledgments section is defined using the "acks" environment
%% (and NOT an unnumbered section). This ensures the proper
%% identification of the section in the article metadata, and the
%% consistent spelling of the heading.
\begin{acks}
This research is supported by the National Research Foundation Singapore and DSO National Laboratories under the AI Singapore Programme (Award Number: AISG2-RP-2020-016). 
The CityU Start-up Grant (No. 9610677) also provides partial support.
Any opinions, findings, conclusions, or recommendations expressed in this material are those of the author(s) and do not reflect the views of the National Research Foundation, Singapore.
We extend our gratitude to all members of the Synteraction Lab and participants for their help in completing this project. We also thank anonymous reviewers for their valuable feedback.
\end{acks}

%%
%% The next two lines define the bibliography style to be used, and
%% the bibliography file.
\bibliographystyle{ACM-Reference-Format}
\bibliography{paper/references}

%%
%% If your work has an appendix, this is the place to put it.
\appendix
% \clearpage
% \newpage

\section{Common Setting}
\label{appendix:common_setting}

\subsection{Scoring}
\label{sec:common_setting:scoring}
Following previous work \cite{weerasinghe_vocabulary_2022, sheshadri_learn_2020, ibrahim_arbis_2018} and our objectives, for each correct recall of an individual L2 word, 1 point was awarded. Variations in the plurality of nouns or verbs, verb tenses, and minor typos in the provided L1 words did not result in a loss of marks (e.g., if the correct L1 word is `reads' and the given L1 word is `read', it was considered correct). A similar scheme was applied to sentence scoring, where each correct identification of an L1 sentence resulted in either 2 or 3 points, depending on the number of unique L2 words in the sentence. 
Partial marks were awarded if the participant recognized the meaning of L2 words and placed them in the correct location within the sentence. The scoring did not consider pronouns, articles, prepositions, and common verbs (i.e., is/are) that appeared in more than one sentence. For example, if the correct L1 sentence is `He reads a book', but the participant provided `He x book', 1 point was awarded. If the participant provided `The book', 0 points were given. If the participant provided, `He read the book', the full score of 2 points was awarded.
For each condition, the total score was calculated by summing the individual scores of words/sentences.

\subsection{Analysis}
\label{sec:common_setting:analysis}
A one-way repeated measures ANOVA (for a single factor with multiple levels) or a factorial ANOVA (for two factors with multiple levels) was used to analyze the quantitative data. When the assumptions of ANOVA were violated, the Friedman test or factorial repeated measures ANOVA after Aligned Rank Transform (ART \cite{wobbrock_aligned_2011}) was employed. The normality of the data was tested using the Shapiro-Wilk test, and sphericity was tested using Mauchly's test. Paired-sample t-tests, ART contrasts\footnote{Using art.con(), \url{https://cran.r-project.org/web/packages/ARTool/vignettes/art-contrasts.html}}, or Wilcoxon signed-rank tests were used as post-hoc tests, with Bonferroni correction applied for multiple comparisons.

\section{\Studyone{}}
\label{appendix:study1}

\begin{figure*}[thbp]
\centering
\begin{subfigure}{\textwidth}
  \centering
  \includegraphics[width=1\linewidth]{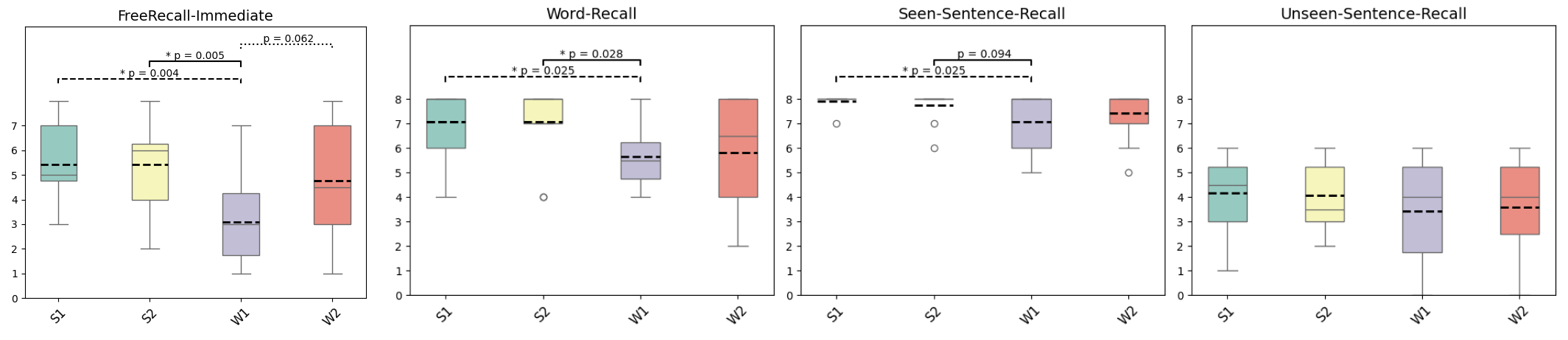}  
  \caption{Immediate \Recall{} scores. The maximum score for each measure is 8, except for \Recall{Unseen-Sentence-}, which has a maximum of 6.}
\end{subfigure}

\begin{subfigure}{\textwidth}
  \centering
  \includegraphics[width=1\linewidth]{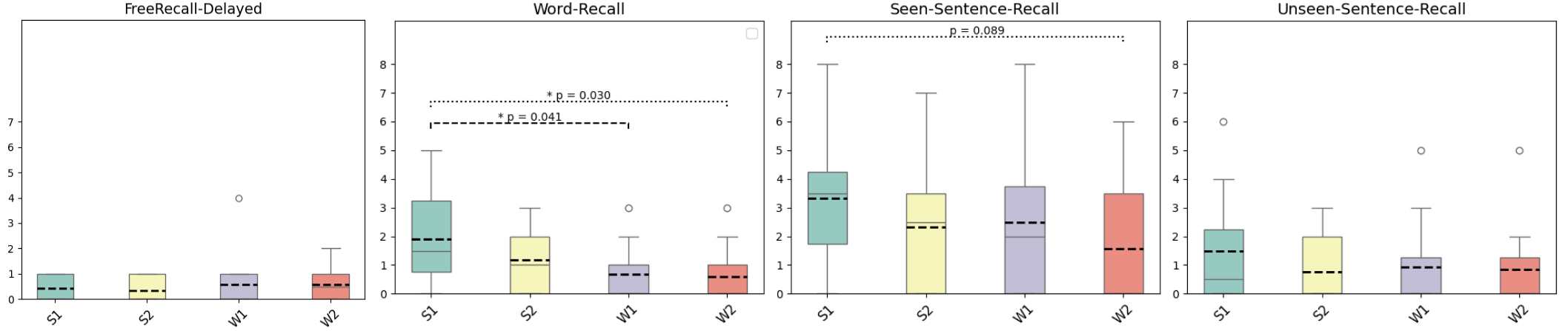}  
  \caption{Delayed (7-day) \Recall{} scores. The maximum score for each measure is 8, except for \Recall{Unseen-Sentence-}, which has a maximum of 6.}
\end{subfigure}

\begin{subfigure}{\textwidth}
  \centering
  \includegraphics[width=1\linewidth]{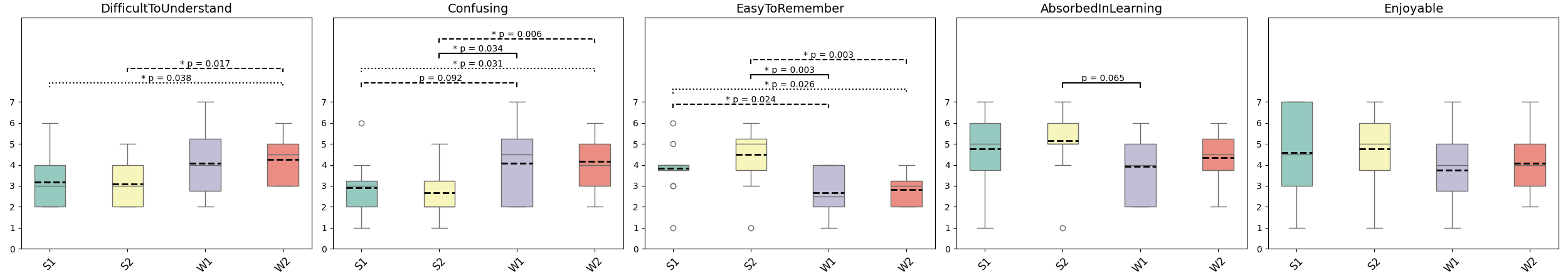}  
  \caption{Subjective ratings on a 7-point Likert scale, where 1 = Strongly Disagree and 7 = Strongly Agree.}
\end{subfigure}

\caption{Measures on recall and perception with 12 participants in \Studyone{}. Here, S = \Sitting{}, W = \Walking{}, 1 = \NoGap{}, and 2 = \WithGap{}. For example, S1 represents, \Sitting{} with \NoGap{}. Dashed lines inside the box plots indicate the mean values.}
% \Description{This figure contains three subfigures showing results from Study 1 with 12 participants. Subfigure (a) displays Immediate Recall scores for four recall types, with maximum scores of 8, except for Unseen-Sentence Recall, which has a maximum of 6. Subfigure (b) presents Delayed Recall scores measured 7 days later, with the same scoring scheme. Subfigure (c) shows participants' subjective ratings on a 7-point Likert scale, evaluating aspects like difficulty, enjoyment, and engagement. Each subfigure uses box plots with dashed lines to indicate mean values. Conditions are labeled S1, S2, W1, and W2, combining Sitting/Walking and NoGap/WithGap presentation styles.}
\label{fig:study1:measures}
\end{figure*}

Table~\ref{tab:study1:measures} and Figure~\ref{fig:study1:measures} indicate the performance of the participants (N=12).

\begin{table*}[hptb]
\caption{Performance and User Ratings with 12 participants in \Studyone{}. Here, S = \Sitting{}, W = \Walking{}, 1 = \NoGap{}, and 2 = \WithGap{}. For example, S1 represents \Sitting{} with \NoGap{}. The \ha{orange} color highlight corresponds to the best (mean) performance for \Sitting{} while the \hb{green} color is for \Walking{}.}
% \Description{This table presents performance measures and user ratings from Study 1 with 12 participants. Conditions are labeled S1, S2, W1, and W2, corresponding to Sitting or Walking combined with NoGap or WithGap content presentation. Measures are grouped into Immediate Recall, Delayed Recall, and User Ratings. Each measure shows the mean (M) and standard deviation (SD). Highlighted cells in orange mark the highest performing Sitting condition per row, while green highlights the highest Walking condition. Measures include recall scores and Likert-scale ratings such as difficulty, confusion, memorability, absorption, and enjoyment.}
\label{tab:study1:measures}
\centering
\begin{tabular}{@{}l|cc|cc|cc|cc@{}}
\toprule
\multicolumn{1}{r}{Condition} &
  \multicolumn{2}{c}{S1} &
  \multicolumn{2}{c}{S2} &
  \multicolumn{2}{c}{W1} &
  \multicolumn{2}{c}{W2} \\
\cmidrule(l){2-9}
\multicolumn{1}{l}{Measure} &
  M & SD & M & SD & M & SD & M & SD \\
\midrule

\multicolumn{9}{l}{\textbf{Immediate Recall Measures}} \\
\Recall{Free-}     & \ha{5.417} & 1.621 & 5.417 & 1.782 & 3.083 & 1.881 & \hb{4.750} & 2.261 \\
\Recall{Word-}                & \ha{7.083} & 1.443 & 7.083 & 1.505 & 5.667 & 1.435 & \hb{5.833} & 2.368 \\
\Recall{Seen-Sentence-}        & \ha{7.917} & 0.289 & 7.750 & 0.622 & 7.083 & 1.165 & \hb{7.417} & 0.996 \\
\Recall{Unseen-Sentence-}      & \ha{4.167} & 1.642 & 4.083 & 1.443 & 3.417 & 2.234 & \hb{3.583} & 2.234 \\

\addlinespace
\multicolumn{9}{l}{\textbf{Delayed Recall Measures}} \\
\Recall{Free-}       & \ha{0.417} & 0.515 & 0.333 & 0.492 & 0.583 & 1.165 & \hb{0.583} & 0.669 \\
\Recall{Word-}                 & \ha{1.917} & 1.730 & 1.167 & 1.193 & \hb{0.667} & 0.985 & 0.583 & 0.996 \\
\Recall{Seen-Sentence-}         & \ha{3.333} & 2.387 & 2.333 & 2.425 & \hb{2.500} & 2.812 & 1.583 & 2.429 \\
\Recall{Unseen-Sentence-}      & \ha{1.500} & 1.977 & 0.750 & 1.138 & \hb{0.917} & 1.621 & 0.833 & 1.528 \\

\addlinespace
\multicolumn{9}{l}{\textbf{User Ratings}} \\
\DifficultToUnderstand{}     & 3.167 & 1.337 & \ha{3.083} & 1.165 & \hb{4.083} & 1.730 & 4.250 & 1.055 \\
\Confusing{}                   & 2.917 & 1.379 & \ha{2.667} & 1.155 & \hb{4.083} & 1.832 & 4.167 & 1.267 \\
\EasyToRemember{}            & 3.833 & 1.193 & \ha{4.500} & 1.508 & 2.667 & 1.155 & \hb{2.833} & 0.835 \\
\AbsorbedInLearning{}        & 4.750 & 1.712 & \ha{5.167} & 1.586 & 3.917 & 1.564 & \hb{4.333} & 1.435 \\
\Enjoyable{}                   & 4.583 & 2.109 & \ha{4.750} & 1.913 & 3.750 & 1.765 & \hb{4.083} & 1.443 \\

\bottomrule
\end{tabular}
\end{table*}

\section{\Pilotfour{}}
\label{appendix:pilot4}

\subsection{Materials}
\label{sec:pilot4:materials}

We selected Norwegian (L2) because its grammar system is straightforward for English speakers to learn, but its vocabulary, derived from Old Norse and influenced by German, is largely distinct from English\footnote{\url{https://thelanguages.com/norwegian/grammar-rules-compared-to-english/}} (see Table~\ref{tab:l1_l2_examples} for examples). This choice was made to avoid confounding factors such as cognates. In cases where close cognates existed, those sentences were excluded from the study (e.g., L1: \quote{book} vs L2: \quote{bok}, L1: \quote{ball} vs L2: \quote{ball}).

Audio pronunciation for the L2 texts was generated using the Google Cloud TextToSpeech API (language: `nb-NO', voice: NEUTRAL, speaking rate: 0.9, format: .mp3) and was manually verified for accuracy.

\subsection{Design and Procedure}
\label{sec:pilot4:design}
The study design was the same as the \Studyone{} (Sec~\ref{sec:study1}).
The procedure was similar to the common setting (Sec~\ref{sec:common:procedure}), with the addition of a pre-test to assess participants' familiarity with the selected L2 words. All participants scored 0 on the pre-test, allowing the post-test scores to be used directly to measure recall accuracy.

\subsection{Results}
\label{sec:pilot4:results}
\begin{figure*}[thbp]
\centering
\begin{subfigure}{\textwidth}
  \centering
  \includegraphics[width=1\linewidth]{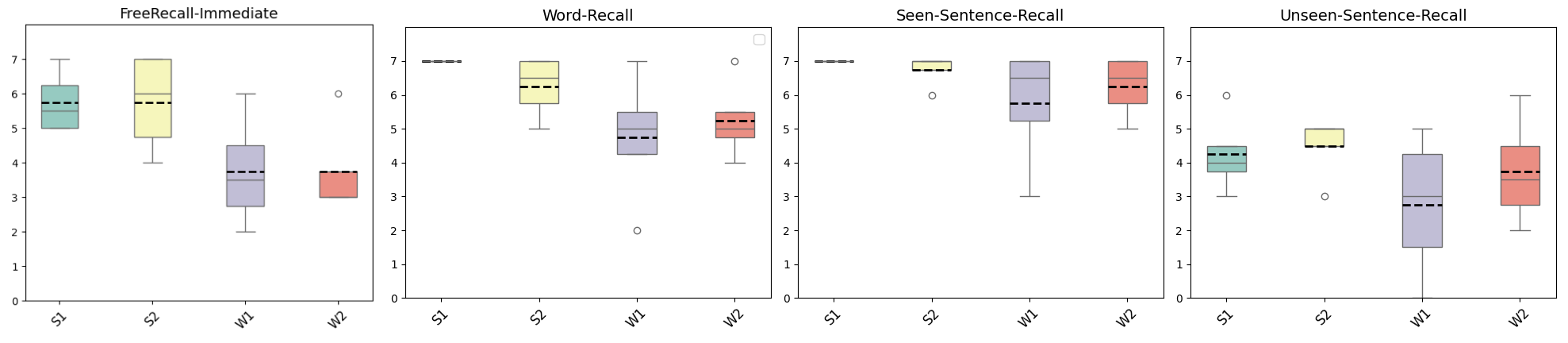}  
  \caption{Immediate \Recall{} scores. The maximum score for each measure is 7, except for \Recall{Unseen-Sentence-}, which has a maximum of 6.}
\end{subfigure}

\begin{subfigure}{\textwidth}
  \centering
  \includegraphics[width=1\linewidth]{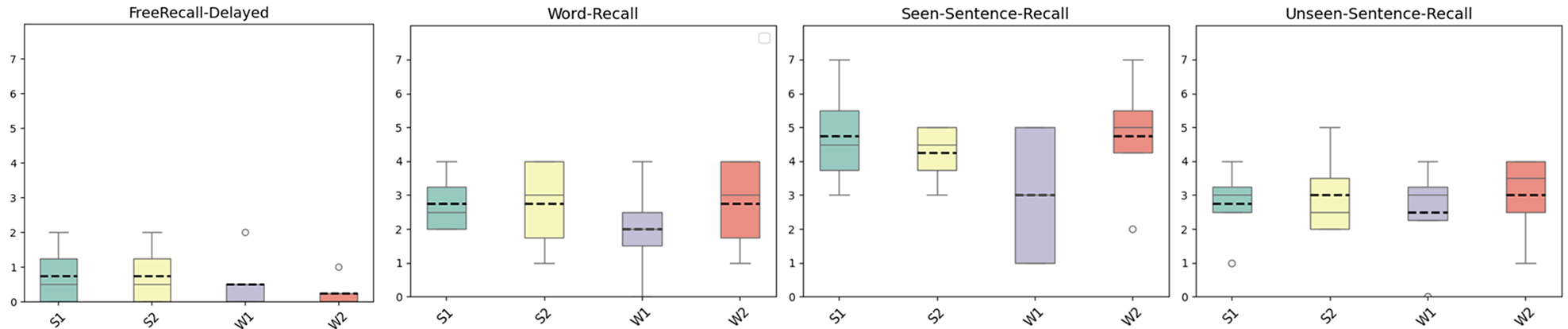}  
  \caption{Delayed (7-day) \Recall{} scores. The maximum score for each measure is 8, except for \Recall{Unseen-Sentence-}, which has a maximum of 6.}
\end{subfigure}

\begin{subfigure}{\textwidth}
  \centering
  \includegraphics[width=1\linewidth]{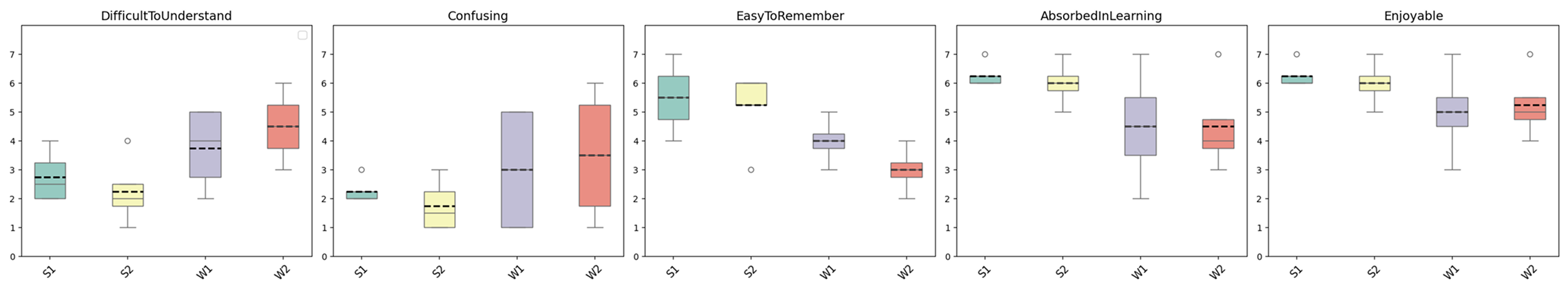}  
  \caption{Subjective ratings on a 7-point Likert scale, where 1 = Strongly Disagree and 7 = Strongly Agree.}
\end{subfigure}

\begin{subfigure}{\textwidth}
  \centering
  \includegraphics[width=0.35\linewidth]{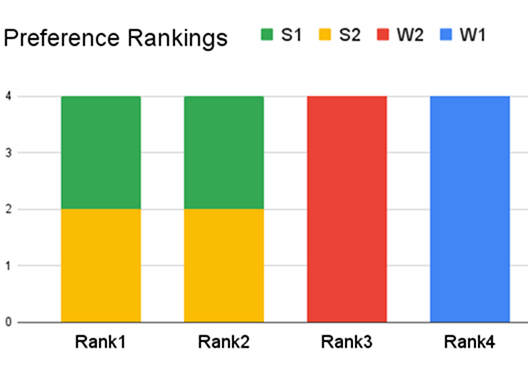}  
  \caption{Preference ranking.}
\end{subfigure}

\caption{Measures on recall and perception with 4 participants in \Pilotfour{}. Here, S = \Sitting{}, W = \Walking{}, 1 = \NoGap{}, and 2 = \WithGap{}. For example, S1 represents, \Sitting{} with \NoGap{}. Dashed lines inside the box plots indicate the mean values.}
% \Description{This figure includes four subfigures reporting data from the Pilot 4 study with four participants. Subfigure (a) shows Immediate Recall scores across different recall types, with a maximum of 7 points per measure, except for Unseen-Sentence Recall (maximum 6). Subfigure (b) presents Delayed Recall scores collected after 7 days, with maximum scores of 8 and 6 for the same measures. Subfigure (c) shows subjective user ratings collected via a 7-point Likert scale, where 1 means Strongly Disagree and 7 means Strongly Agree. These ratings assess perceived difficulty, confusion, ease of remembering, absorption, and enjoyment. Subfigure (d) displays participants’ preference rankings for the four study conditions (S1, S2, W1, W2). All box plots include dashed lines to denote mean values.}
\label{fig:pilot4:measures}
\vspace*{-5mm}
\end{figure*}

Table~\ref{tab:pilot4:measures} and Figure~\ref{fig:pilot4:measures} indicate the performance of the participants (N=4).

\begin{table*}[hptb]
\caption{Performance and User Ratings with 4 participants in \Pilotfour{}. Here, S = \Sitting{}, W = \Walking{}, 1 = \NoGap{}, and 2 = \WithGap{}. For example, S1 represents, \Sitting{} with \NoGap{}. The \ha{orange} color highlight corresponds to the best (mean) performance for \Sitting{} while the \hb{green} color is for \Walking{}.}
% \Description{This table shows performance and user ratings from the Pilot 4 study with 4 participants. Results are presented across four conditions: S1, S2, W1, and W2, combining Sitting or Walking with NoGap or WithGap presentation styles. Measures are grouped into Immediate Recall, Delayed Recall, and User Ratings. Each row contains mean (M) and standard deviation (SD) values. Orange highlights mark the highest-performing Sitting condition for each measure; green highlights indicate the highest-performing Walking condition. Recall metrics include word and sentence recall tasks. User ratings are based on Likert-scale items about difficulty, confusion, memory ease, engagement, and enjoyment.}
\label{tab:pilot4:measures}
\centering
\begin{tabular}{@{}l|cc|cc|cc|cc@{}}
\toprule
\multicolumn{1}{r}{Condition} &
  \multicolumn{2}{c}{S1} &
  \multicolumn{2}{c}{S2} &
  \multicolumn{2}{c}{W1} &
  \multicolumn{2}{c}{W2} \\
\cmidrule(l){2-9}
\multicolumn{1}{l}{Measure} &
  M & SD & M & SD & M & SD & M & SD \\
\midrule

\multicolumn{9}{l}{\textbf{Immediate Recall Measures}} \\
\Recall{Free-}     & \ha{5.750} & 0.957 & 5.750 & 1.500 & 3.750 & 1.708 & \hb{3.750} & 1.500 \\
\Recall{Word-}     & \ha{7.000} & 0.000 & 6.250 & 0.957 & 4.750 & 2.062 & \hb{5.250} & 1.258 \\
\Recall{Seen-Sentence-} & \ha{7.000} & 0.000 & 6.750 & 0.500 & 5.750 & 1.893 & \hb{6.250} & 0.957 \\
\Recall{Unseen-Sentence-} & 4.250 & 1.258 & \ha{4.500} & 1.000 & 2.750 & 2.217 & \hb{3.750} & 1.708 \\

\addlinespace
\multicolumn{9}{l}{\textbf{Delayed Recall Measures}} \\
\Recall{Free-}     & 0.750 & 0.957 & 0.750 & 0.957 & \hb{0.500} & 1.000 & 0.250 & 0.500 \\
\Recall{Word-}     & \ha{2.750} & 0.957 & 2.750 & 1.500 & 2.000 & 1.633 & \hb{2.750} & 1.500 \\
\Recall{Seen-Sentence-} & \ha{4.750} & 1.708 & 4.250 & 0.957 & 3.000 & 2.309 & \hb{4.750} & 2.062 \\
\Recall{Unseen-Sentence-} & 2.750 & 1.258 & \ha{3.000} & 1.414 & 2.500 & 1.732 & \hb{3.000} & 1.414 \\

\addlinespace
\multicolumn{9}{l}{\textbf{User Ratings}} \\
\DifficultToUnderstand{} & 2.750 & 0.957 & \ha{2.250} & 1.258 & \hb{3.750} & 1.500 & 4.500 & 1.291 \\
\Confusing{}             & 2.250 & 0.500 & \ha{1.750} & 0.957 & 3.000 & \hb{2.309} & 3.500 & 2.380 \\
\EasyToRemember{}        & \ha{5.500} & 1.291 & 5.250 & 1.500 & \hb{4.000} & 0.816 & 3.000 & 0.816 \\
\AbsorbedInLearning{}    & \ha{6.250} & 0.500 & 6.000 & 0.816 & 4.500 & 2.082 & \hb{4.500} & 1.732 \\
\Enjoyable{}             & \ha{6.250} & 0.500 & 6.000 & 0.816 & 5.000 & 1.633 & \hb{5.250} & 1.258 \\

\bottomrule
\end{tabular}
\end{table*}

\end{document}